\documentclass{SCAE}

\usepackage{graphicx}
\usepackage[boxed,linesnumbered]{algorithm2e}
\usepackage{subfigure}
\usepackage{amsmath}
\usepackage{float}
\usepackage{verbatim}

\def\leq{\leqslant}
\def\geq{\geqslant}
       
  \def\tlj{\end{document}}  \newsymbol\wjzhml 203F

\begin{document}


\Year{2010} 
\Month{January}
\Vol{53} 
\No{1} 
\BeginPage{1} 
\EndPage{18} 
\AuthorMark{{\rm Li K,} et al.}
\DOI{} 

\title{A Fractal and Scale-free Model of Complex Networks with Hub Attraction Behaviors}{}


\author[1]{Li Kuang}{}
\author[2]{Bojin Zheng}{Corresponding author (email: zhengbojin@gmail.com)}
\author[3]{Deyi Li}{}
\author[1]{Yuanxiang Li}{}
\author[4]{Yu Sun}{}

\address[{\rm1}]{State Key Laboratory of Software Engineering, Computer School, Wuhan University\\ Wuhan {\rm 430072}, China,}
\address[{\rm2}]{College of Computer Science, South-Central University For Nationalities\\ Wuhan {\rm 430074}, China,}
\address[{\rm3}]{School of Software, Tsinghua University\\Beijing {\rm 100084}, China,}
\address[{\rm4}]{School of Computer and Electronics and Information, Guangxi University\\ Nanning {\rm 530004} , China,}
\maketitle

\vspace{-2mm} \begin{center}\bahao{Received August 22, 2008;
accepted June 6, 2009}
\end{center}

\vspace{2mm}\hspace{7.5mm}\begin{tabular}{p{0.88\textwidth}}\hline\\
\end{tabular}\vspace{-8mm}
\begin{center}
\parbox{14.5cm}{\begin{abstract}
It is widely believed that fractality of complex networks origins from hub repulsion behaviors (anticorrelation or disassortativity), which means large degree nodes tend to connect with small degree nodes. This  hypothesis was demonstrated by a dynamical growth model, which evolves as the inverse renormalization procedure  proposed by Song \textit{et al}. Now we find that the dynamical growth model is based on the assumption that all the cross-boxes links has the same probability $e$ to link to the most connected nodes inside each box. Therefore, we modify the growth model by adopting the flexible probability $e$, which makes hubs  have higher probability to connect with hubs  than non-hubs. With this model, we find some fractal and scale-free  networks have hub attraction behaviors (correlation or assortativity). The results are the counter-examples of former beliefs.

\end{abstract}}\end{center}

\begin{center}
\parbox{14.5cm}{
}\end{center}

\renewcommand{\baselinestretch}{1.2}{\scriptsize
\hspace{7.5mm}\begin{tabular}{p{0.88\textwidth}}
\hline\vspace{0.01mm}
 {\bf Citation}\quad~
\\
\hline
\end{tabular}}

\renewcommand{\baselinestretch}{1.09} \baselineskip 12pt\parindent=10.8pt
\wuhao \vspace{5mm}
\section{Introduction}
 Fractal Geometry theory is proposed by Mandelbrot to explain the  mathematical set which has a fractal dimension exceeds its topological dimension[1,2]. Fractals are patterns that exhibit self-similar at different length scales. Nowadays, more and more researchers focus on the study of complex networks. A critical property of complex networks is the  degrees of nodes follow  power law distribution[3,4,5]. The probability \textit{P(k)} of number of connections $k$ to a node fulfill  power-law relation:
\begin{equation}
 P(k)\approx {k}^{-\gamma }.
 \label{degreedistribution}
 \end{equation}

  As  fractal geometry focus on  mathematical set on Euclid space, some researchers  wondered whether complex networks in topological space have fractality property [9,10].  Among them, Song \textit{et al}'s researches of fractality property in complex networks have attracted extensive attentions. They proposed a renormalization procedure implemented by box covering method[6], which was inspired from `box counting' method applied in Euclid space, to tile the networks into boxes with a given box size as shown in Fig.1. The box size ${\ell}_{B}$  is the upper bound of shortest path between nodes in each box. The process is iterated until only  single node left. Through the procedure, they found that many real world networks, such as the World Wide Web, Protein-Protein Interaction networks, and Cellular networks have scale-invariance and fractality properties. The scale-invariance means the degree distribution of renormalized networks still follow the power law under different length scale renormalizations, and the fractality means the number of boxes ${N}_{B}$, which is needed to cover the whole network, is approximate to power $-{d}_{B}$ of the box size ${\ell }_{B}$. This can be  defined as equation(\ref{lbnb}):
 \begin{equation}
 {N}_{B}({\ell }_{B})\approx{ {\ell }_{B}}^{-{d}_{B}}.
 \label{lbnb}
 \end{equation}

\begin{figure}

  \centering

 \includegraphics[width=1\textwidth ]{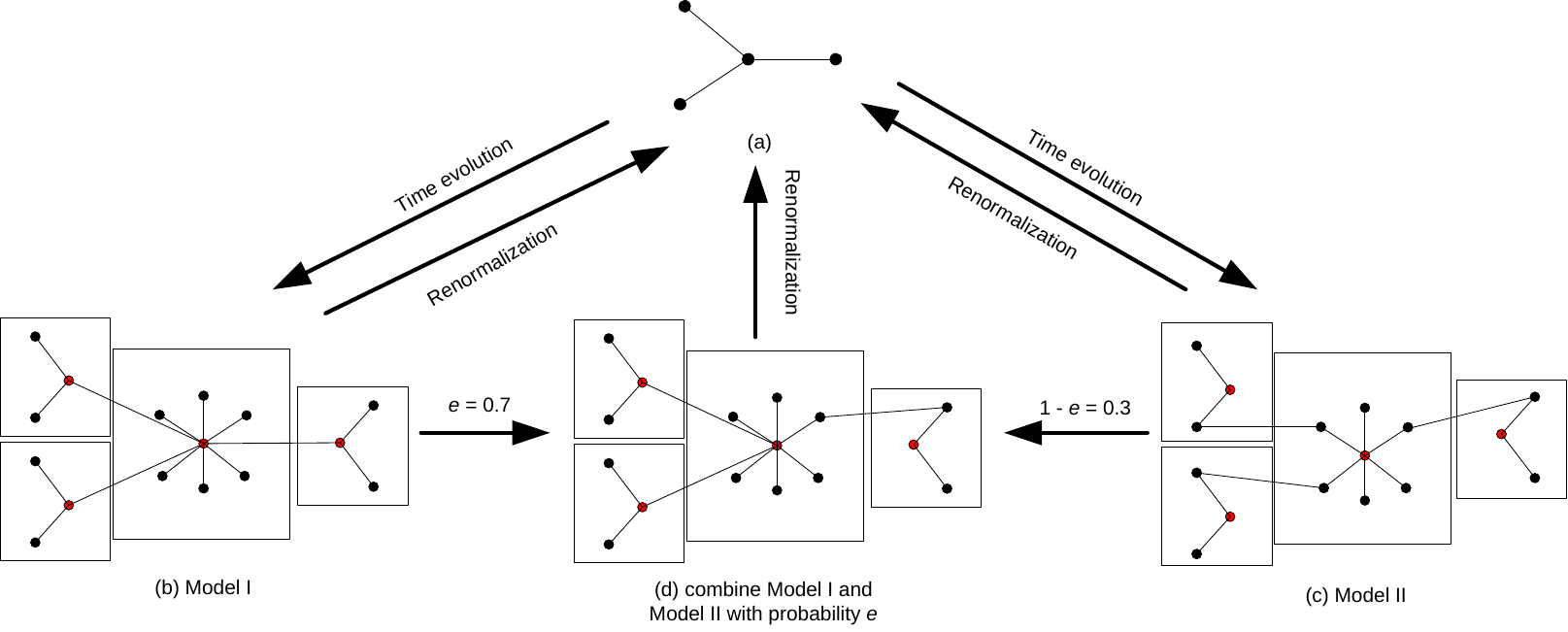}
  \caption{Network dynamical growth with parameters: degree multiplier \textit{m} = 2 and box size ${\ell}_{B}=3$. The dynamical growth process can be seen as the inverse renormalization procedure. \textbf{(a)}, Four nodes at \textit{t} = 0 that could growth to networks in (b), (c) and (d) through time evolution. \textbf{(b)},  Mode I alone, where most connected nodes in each box connect directly to each other. \textbf{(c)}, Mode II alone, where boxes connected by less connected node. \textbf{(d)}, Combination of Mode I and II with probability \textit{e }= 0.7.  }
   \label{originmodel}

\end{figure}

Song \textit{et al} analyzed the origin of the fractality of complex networks later[7]. They proposed the dynamical growth model (DMG) from the perspective of networks evolution based on the ideas of Barab\'{a}si-Albert model[4]. It is the inverse renormalization procedure as showed in Fig.1. The network evolves as time step \textit{t} increases. Every node at time step $t-1$  with degree $k(t-1)$ will evolve to a visual box with $mk(t-1)$ new nodes (black nodes)  generated inside each box. And initial nodes (red nodes) connect to all the new nodes in each box. They proposed two kinds of models with difference on generating cross-boxes links. In  Model I, most connected nodes in each box connect directly to each other as shown in Fig.1(b), while in Model II, links between boxes connect to less connected nodes inside each boxes as shown in Fig.1(c).  The extensive growth model is the combination of Model I and Model II with a stationary probability \textit{e} through whole network as showed in Fig.1(d), where \textit{e} is defined as the measurement of the level of `hub' attraction.
Their results have shown networks with  higher probability of \textit{e} tend to be non-fractal, yet networks with lower probability of \textit{e} inclined to be fractal. Therefore, they concluded that `hub' repulsion is the cause of fractality.

Similar research found that fractal scale-free networks are disassortative mixing[14].  There was also research work illustrated the fractal networks have to fulfill the criticality condition: the skeleton, which has branching tree structure, grows from root node (most connected node in network) perpetually with offspring neither flourishing nor dying out[16].

However, the above assertions are based on experiments rather than theoretical proof. In this paper, we get different results by simply modifying the DGM model.
As we know, hubs are defined as the most connected nodes in the whole network[3]. We notice Song \textit{et al}  presumed the most connected nodes in each box as the hubs of whole network. In fact, due to the power law degree distribution, most of the highest degree nodes in boxes have much lower degree compared with the real hubs in the network. Therefore, we apply  flexible probability $e$ mechanism to make real hubs have higher probability to connect with each other, while lower the probability of connection between non-hub nodes. By applying this mechanism, we can achieve fractal and scale-free networks (HADGM) with strong hub attraction. Moreover, we find relative researches that supports our statement. Some optimization networks exhibit both fractality and assortativity mixing properties[18]. Our research gives a fundamental challenge to the former researches on the origin of the fractal complex networks. We notice the fractality has strong correlation with the diameter of networks. As long as the diameter grows exponential with time evolution, the networks will preserve fractality property.

This paper is organized as follows. In Sec.2, we introduce the HAGMD model. The subsections present the flexible probability $e$  mechanism and inside box link-growth method. In Sec.3, we analyze the properties of  HAGMD model, such as fractality, scale-free and correlations. In Sec.4, we introduce the fractal optimization network and compare the assortativity of  all the models. At last, In Sec.5, we summery our works and discuss the origin of  fractality.


\section{Hub attraction dynamical growth model}
As shown in Fig.2, based on the dynamical growth framework of DGM model, we make probability $e$ flexible (hubs have higher probability to connect to each other). This mechanism alone could make hubs connected together, while persevering the fractality property. However, we find DGM networks are only spinning trees without loops, which are not similar with real networks. Therefore, we also propose the inside box link-growth method, which add the correlation between high degree nodes.

 \begin{figure*}[h]
\begin{minipage}[t]{0.34\linewidth}
\centering
\subfigure[]{
\includegraphics[width=2.2in]{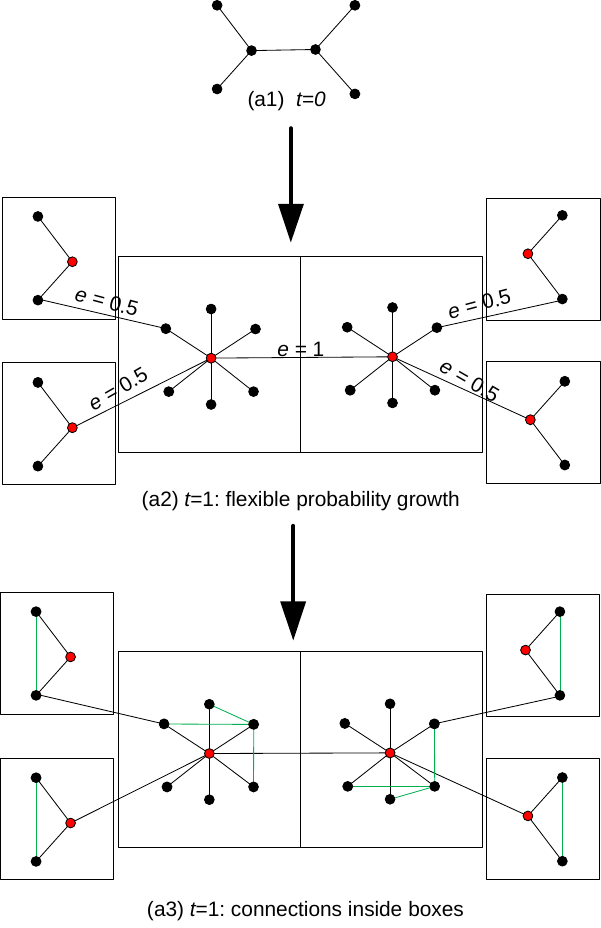}}

\label{fig:side:a}
\end{minipage}%
\begin{minipage}[t]{0.7\linewidth}
\centering
\subfigure[]{
\includegraphics[width=3.5in]{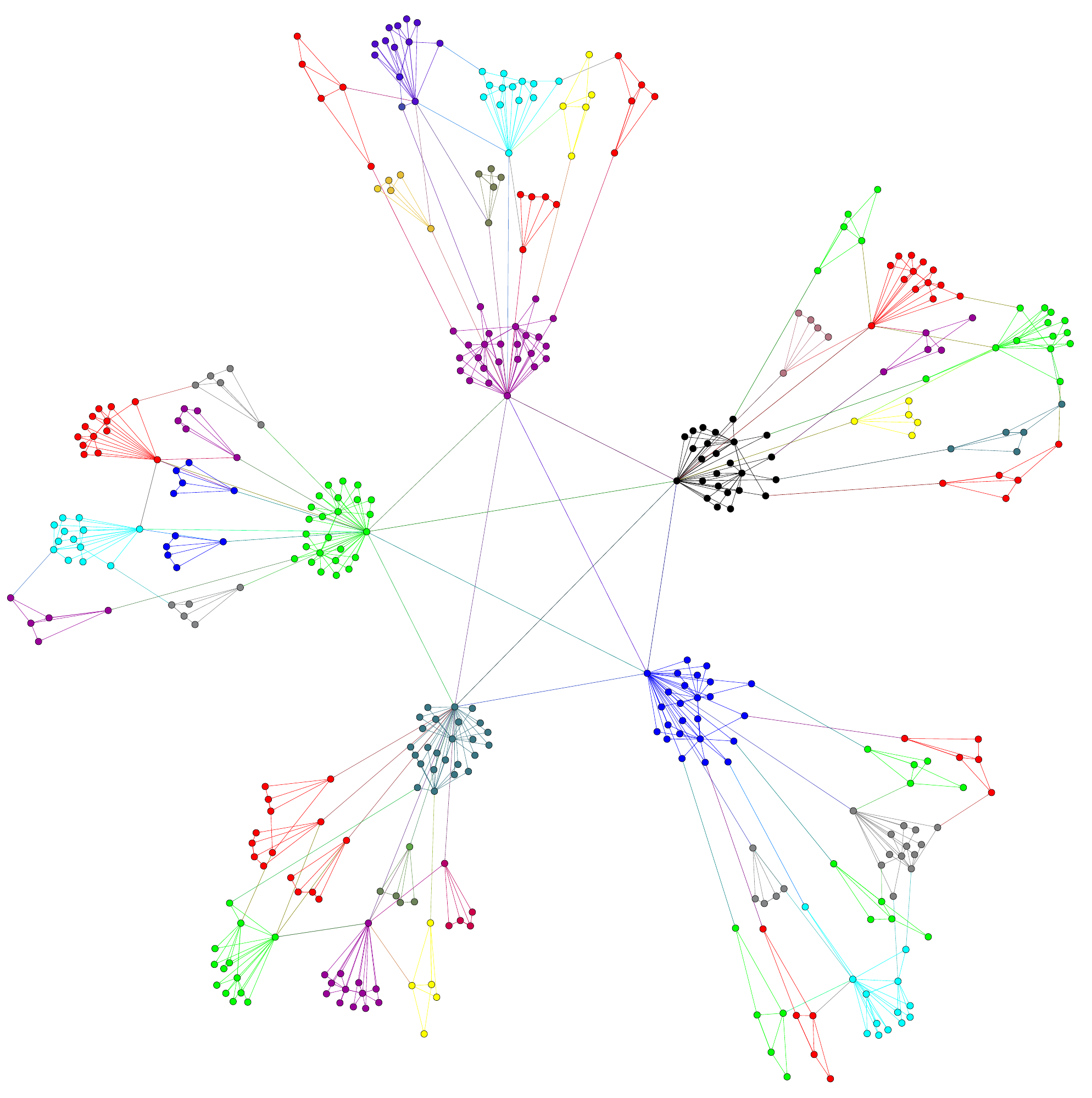}}

\label{fig:side:b}
\end{minipage}
\caption{Hub attraction dynamical growth model. \textbf{(a1)}, Five nodes at time step $t = 0$. \textbf{(a2)} Dynamical growth network at $t = 1$ with flexible probability $e$, which $e=1$ between high degree nodes, and $e=0.5$ between low degree nodes. \textbf{(a3)} The method of generating connections inside box (green line), where we random pick one new generated nodes and link it to other  $\tilde{k}(t-1)$  new nodes.\textbf{ (b)} HADGM network at time step $t=2$ which have initial 5 nodes at $t=0$ .}
\label{model}
\end{figure*}

\subsection{Flexible probability \textit{e} mechanism}

 In the dynamical growth processes, each edge at time step $t-1$ will become cross-boxes link at time step $t$. We define  probability \textit{e} as piecewise function depending on the degrees of nodes (${k}_{1}(t-1)$ and ${k}_{2}(t-1)$) attached on both side of edge at time step $t-1$. As shown in equation(\ref{probe}):
 \begin{equation}
 e({k}_{1}(t-1),{k}_{2}(t-1))=\begin{cases}
a & \text{ if } \frac{{k}_{1}(t-1)}{{k}_{max}(t-1)}> T \: \text{ and } \: \frac{{k}_{2}(t-1)}{{k}_{max}(t-1)}> T  \\
b & \text{ if } \frac{{k}_{1}(t-1)}{{k}_{max}(t-1)}\leq T \:\:\: \text{ or} \:\:\:\: \frac{{k}_{2}(t-1)}{{k}_{max}(t-1)}\leq T\end{cases} \\.
\label{probe}
\end{equation}
where $T,a,b$ are predefined parameters with domain as $0\leq b < a\leq 1 $ , $ 0<T\leq 1$, and ${k}_{max}(t-1)$ is the max degree in network at time step $t-1$. Therefore, hubs in the network have higher probability to connect to each other, we can even define $a=1$ to let all hubs connected with each other as shown in Fig.2(a2) and Fig.2(b), where $T=0.5$ and $b=0.5$.

\subsection{Inside box link-growth method}
As shown in Fig.2(a3), in each time step of dynamical growth, we apply inside box link-growth method after the nodes growth phase.   In time step $t$, we add $\tilde{k}(t-1)$ links in each box. Therefore, we add $2\tilde{K}(t-1)$ links in the whole network, where $\tilde{K}(t)$ is the total number of links  at time \textit{t}.
The  method does not affect the fractality and scale-free properties of the networks, but it increase the level of correlation of the networks. We will prove them in the following paragraph.

 \begin{figure*}[h]
\begin{minipage}[t]{0.5\linewidth}
\centering
\subfigure[]{
\includegraphics[width=2.8in]{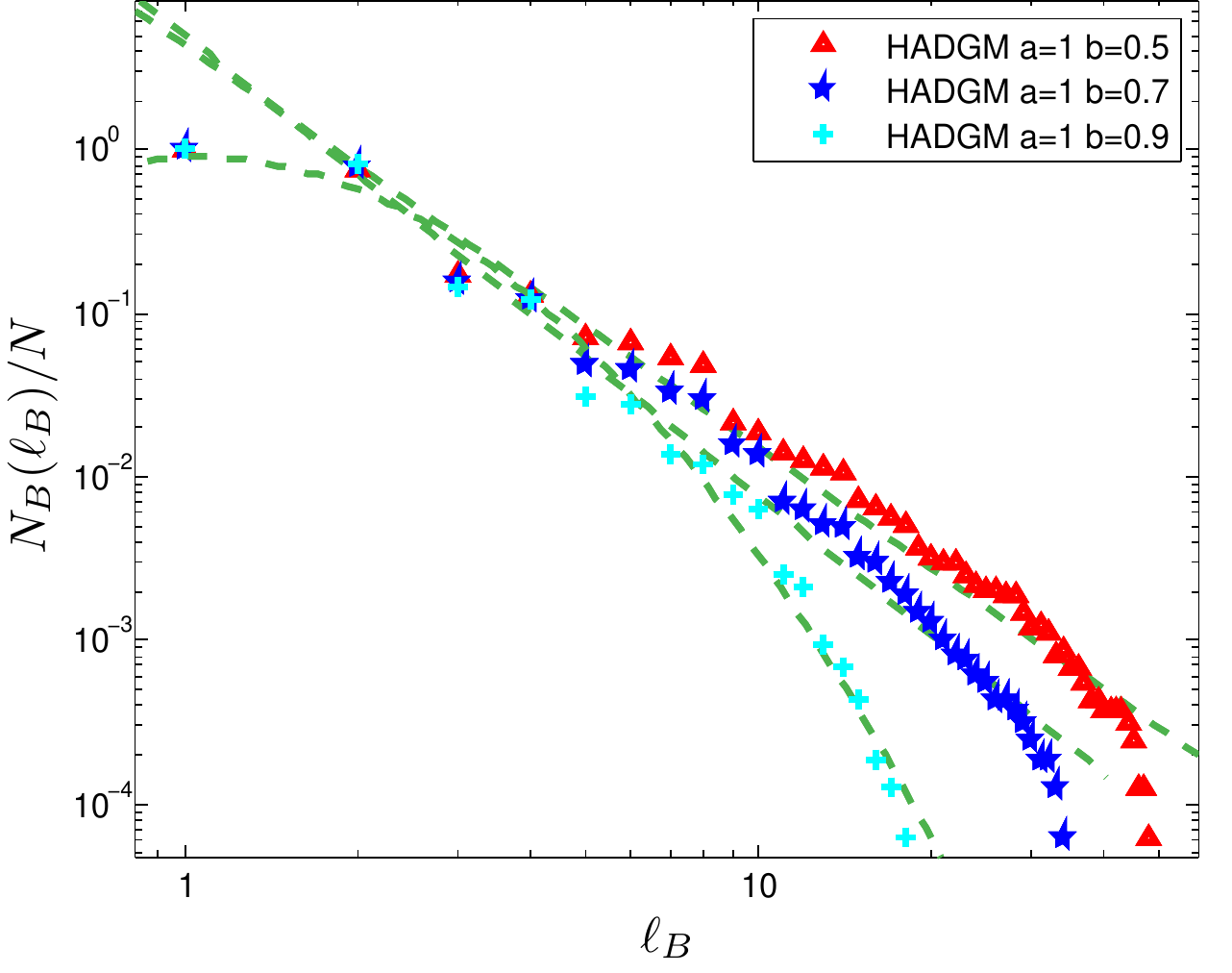}}
\end{minipage}%
\begin{minipage}[t]{0.5\linewidth}
\centering
\subfigure[]{
\includegraphics[width=2.8in]{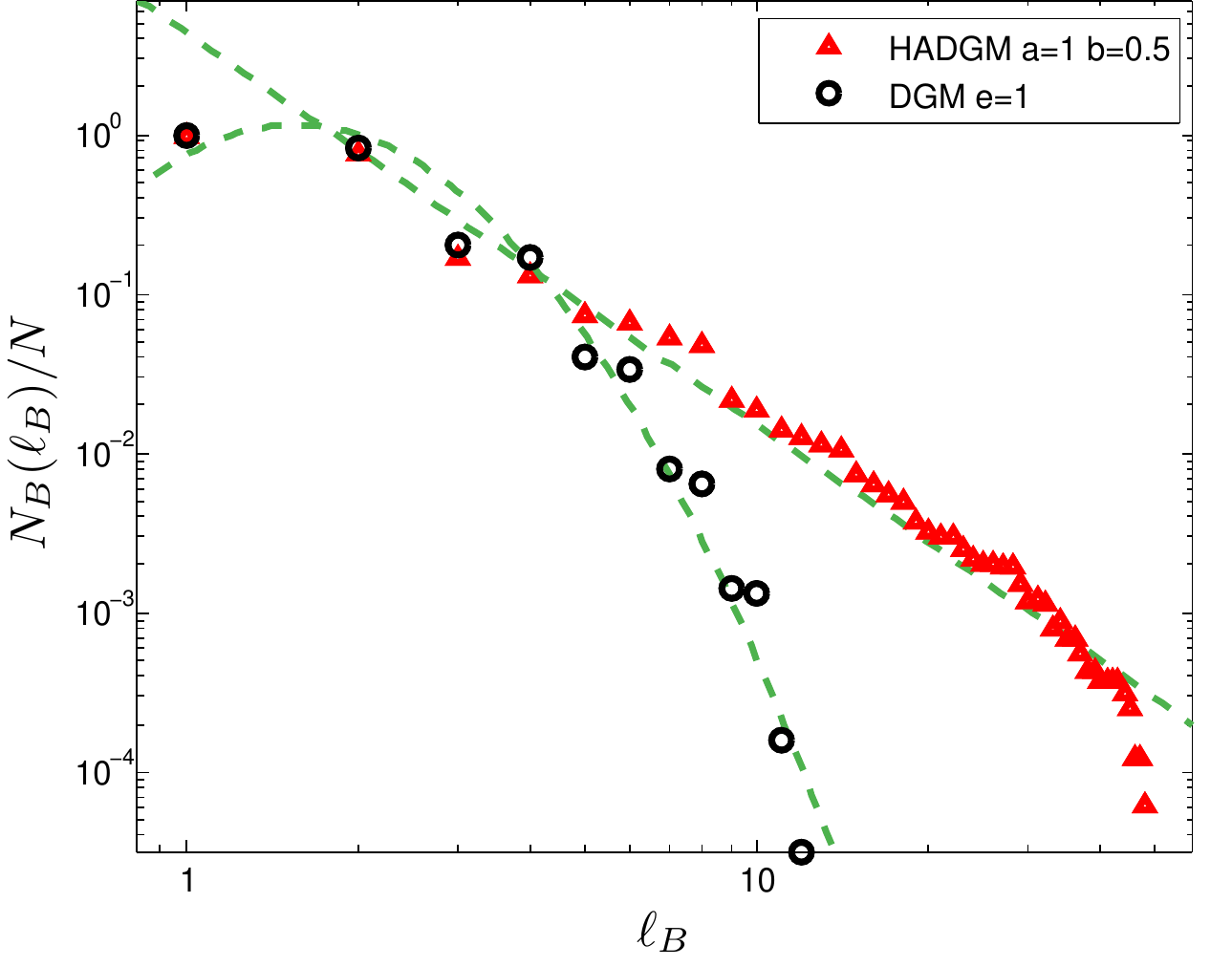}}
\end{minipage}

\begin{minipage}[t]{1\linewidth}
\centering
\subfigure[]{
\includegraphics[width=2.8in]{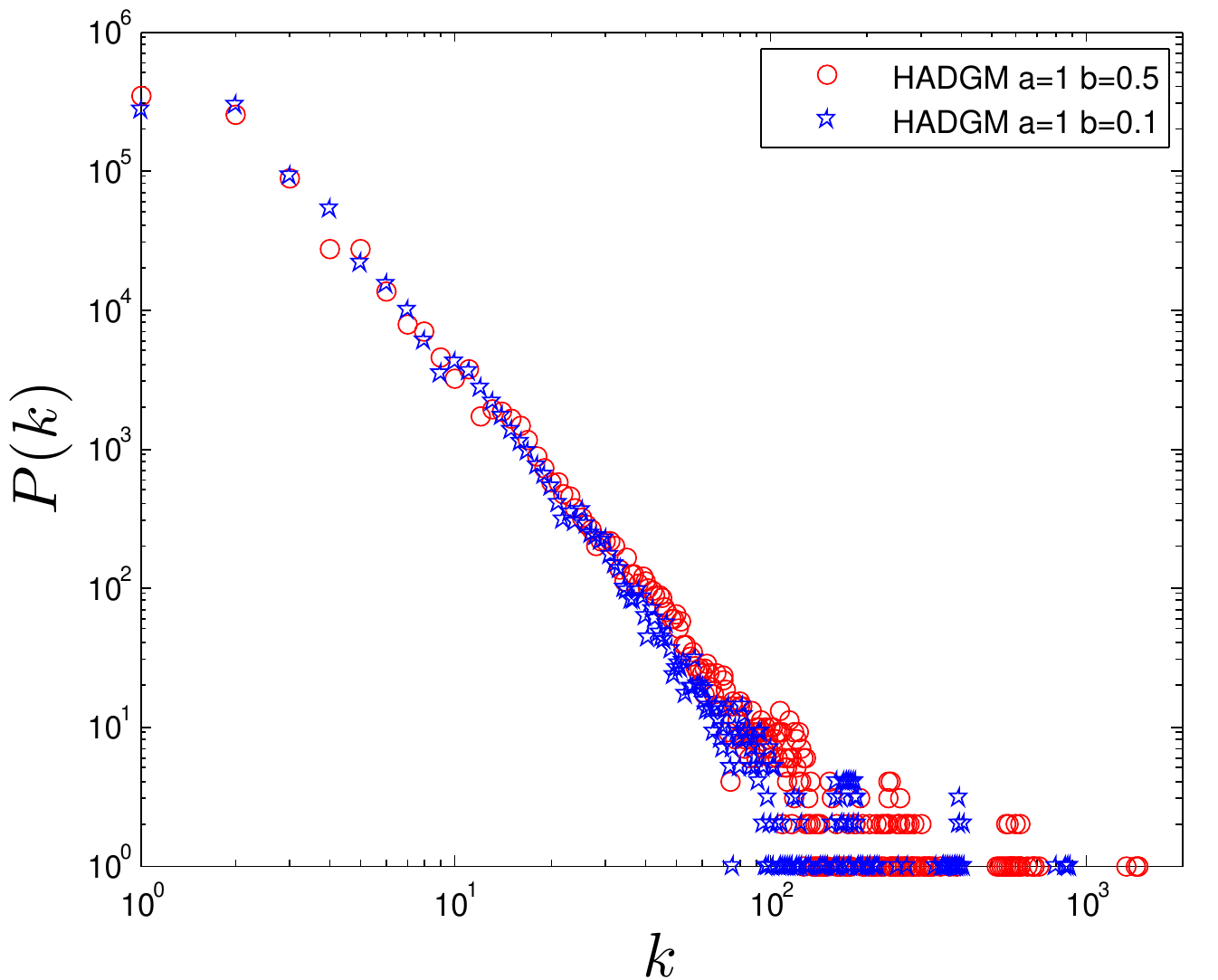}}
\end{minipage}%

\caption{\textbf{(a)}, Log-Log plot of ${N}_{B}({\ell }_{B})$ vs ${\ell }_{B}$ of HADGM networks with difference on parameter $b$. HADGM networks shows fractal property with $b=0.5$ and $b=0.7$, as $b$ increase to $0.9$ the network becomes non-fractal. \textbf{(b)}, Comparison of ${N}_{B}({\ell }_{B})$ vs ${\ell }_{B}$ of HADGM network (fractal with $a=1$, $b=0.5$) and DGM network (non-fractal with $e=1$). \textbf{(c)}, Degree distribution of HADGM networks with same parameters $a=1$, $T=0.5$ and $m=2$, while have difference on $b=0.5$ (red circle) and $b=0.1$ (blue star). }
\label{branchingratio}
\end{figure*}

\section{Properties of HADGM}
In order to measure the properties of HADGM networks, we apply mathematic framework as equation(\ref{nodedeg}) (Further deduction see the Appendix).
 \begin{equation}
 \begin{split}
\tilde{N}(t)&\approx (2m+3)\tilde{N}(t-1) \;\;  for \;\;  t>1,\\
\tilde{k}(t)&= (m+\bar{e})\tilde{k}(t-1),\\
\tilde{L}(t)&= (3-2\bar{e})\tilde{L}(t-1)+2e,
 \end{split}
\label{nodedeg}
 \end{equation}
 where $\tilde{N}(t)$  is the total number of nodes at time step $t$, $\tilde{k}(t)$ is the maximum degree inside a box at time step $t$, $\tilde{L}(t)$ is the diameter of the network at time step $t$, and $\bar{e}$ is the average of flexible probability $e$.

\subsubsection{Fractality of HADGM}
The networks evolve dynamically as the inverse renormalization procedure. Each node at time step $t-1$ will grow into a  virtual box at time step $t$, thus based on equation(\ref{nodedeg}) we have $\tilde{N}({t}_{1})/\tilde{N}({t}_{2})={N}_{B}({\ell}_{B})/N=(2m+3)^{{t}_{1}-{t}_{2}}$. And we also have $(\tilde{L}({t}_{2})+{L}_{0})/(\tilde{L}({t}_{1})+{L}_{0})={\ell}_{B}+{L}_{0}=(3-2\bar{e})^{{t}_{2}-{t}_{1}}$, where ${L}_{0}$ is the initial diameter[7]. Therefore, by inputting equation({\ref{lbnb}}) and replacing the time interval ${{t}_{2}-{t}_{1}}$, we have fractal dimension ${d}_{B}$ as:
 \begin{equation}
 {d}_{B}=\ln(2m+3)/\ln(3-2\bar{e}).
\label{fradimension}
 \end{equation}
Due to the power law degree distribution, for $T\geq 0.5$ the average of probability $\bar{e}$ is close to value $b$. Thus, if we choose low value of $b$, the average probability $\bar{e}$ is obviously lower than $1$. So we will have finite fractal dimension ${d}_{B}$, which imply the fractality of HADGM. As shown in Fig.3(a), we apply box covering algorithm[8,11] to HADGM networks, which have different value $b$  and other parameters are the same. Thus, we find  the HADGM networks are fractal if $b<0.9$. We notice when $b$ increase to a certain point $b=0.9$, then $\bar{e}\rightarrow 1$ and ${d}_{B}\rightarrow \infty$, so the HADGM network become non-fractal.
\begin{figure*}[h]
\begin{minipage}[t]{0.5\linewidth}
\centering
\subfigure[]{
\includegraphics[width=3.2in]{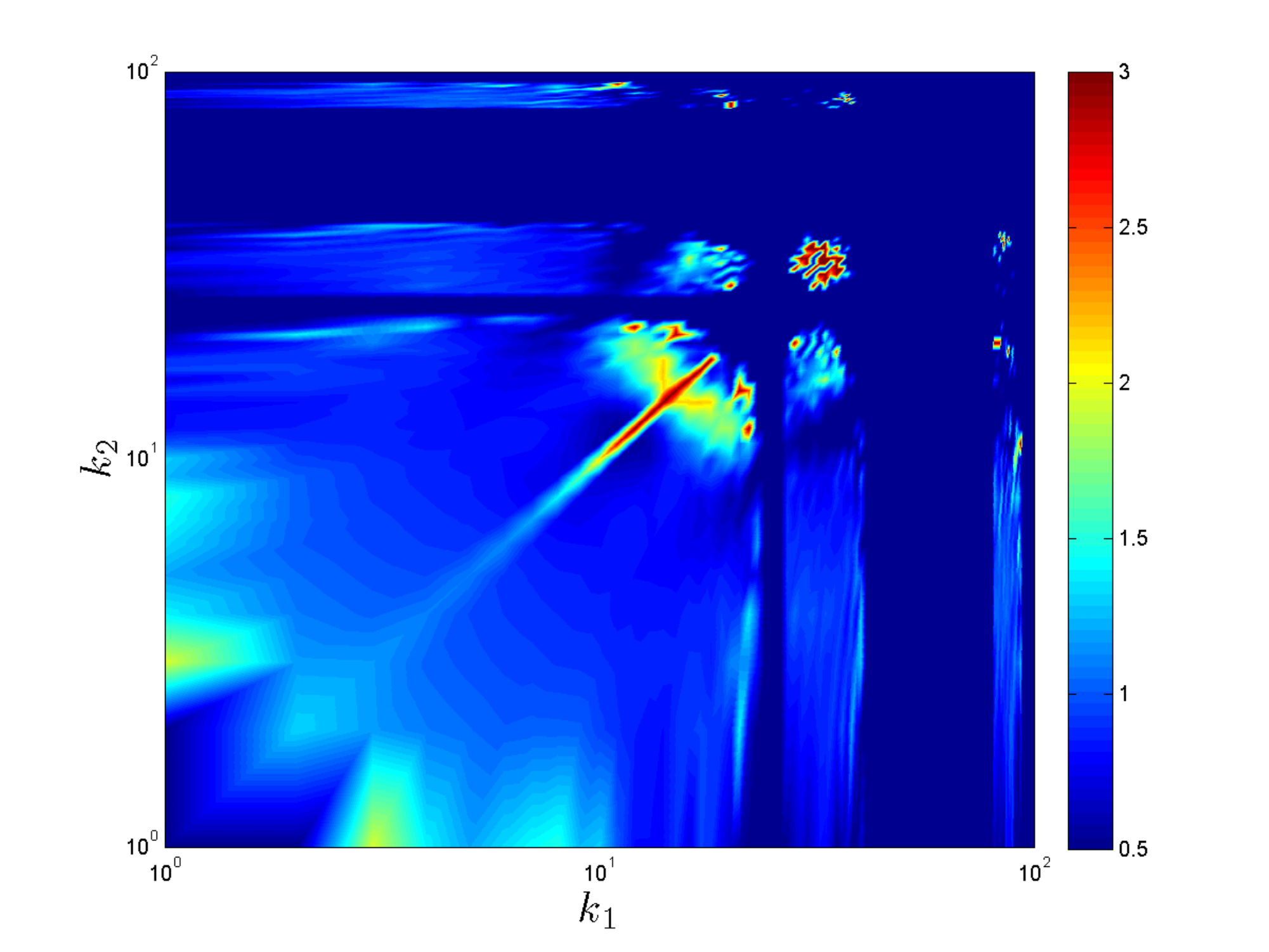}}

\label{fig:side:a}
\end{minipage}%
\begin{minipage}[t]{0.5\linewidth}
\centering
\subfigure[]{
\includegraphics[width=3.2in]{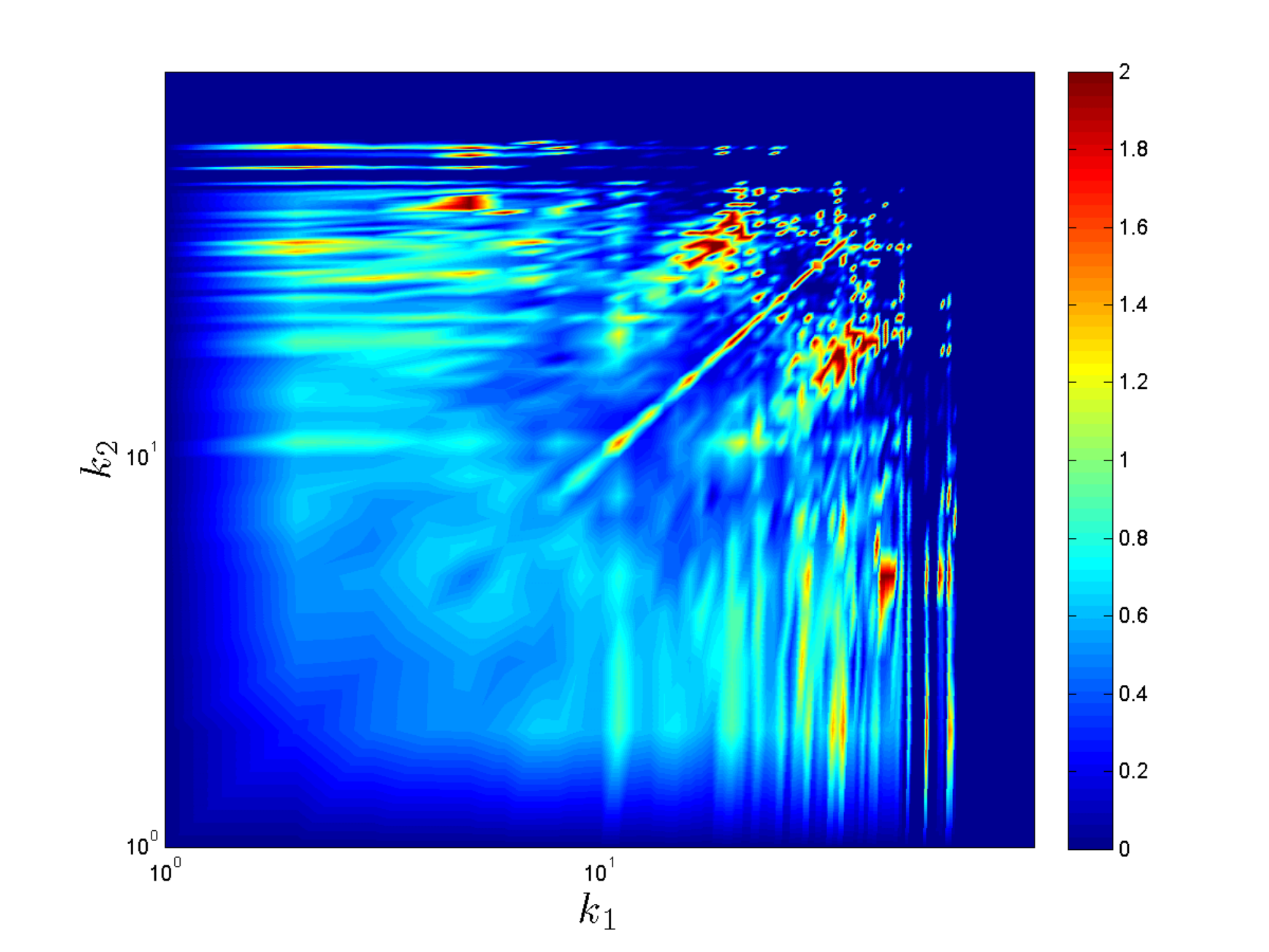}}
\end{minipage}
\caption{ Comparison of Correlation : Different colors relate to different strength of correlation. Red color in the plot represent strong correlation. HADGM network shows stronger correlation between high degree nodes than other two networks. \textbf{(a)}， Plot of ${R}_{HADGM}({k}_{1},{k}_{2})/{R}_{DGM}({k}_{1},{k}_{2})$ that compared correlation of HADGM network (fractal with $a=1$, $b=0.5$)  with DGM network (non-fractal with $e=1$) \textbf{(b)}， Plot of ${R}_{HADGM}({k}_{1},{k}_{2})/{R}_{WWW}({k}_{1},{k}_{2})$ that compared correlation of HADGM network (fractal with $a=1$, $b=0.5$)  with the World Wide Web}
\label{correlation}
\end{figure*}

\subsubsection{Scale-free of HADGM}
The recursive growth of $\tilde{k}(t)$ lead to the power law degree distribution of HADGM networks as shown in equation(\ref{nodedeg}). The results are illustrated in Fig.3(c), the  HADGM networks with different parameter $b$ show clear power law degree distribution with fat tail.  And the two HADGM networks have differences on slope of the power law curve due to the difference on average probability $\bar{e}$.

\subsubsection{Correlation and Fractality}

Former research[7] asserted that fractal networks tend to show anticorrelation behaviors, while non-fractal networks exhibit correlation behaviors. The results are achieved by comparing the strength of anticorrelation between different networks. Such as non-fractal DGM model with $e=1$ has stronger correlation than fractal DGM model with $e=0.8$.   However, our comparison results show the fractal HADGM model has even stronger correlation than non-fractal DGM model with $e=1$. Therefore, we get different results with the assertion that the hub repulsion behaviors give rise to fractality property.

The correlation is quantified by equation: $R({k}_{1},{k}_{2})=P({k}_{1},{k}_{2})/P_{r}({k}_{1},{k}_{2})$, which illustrates the correlation topological properties of a network[17]. Given a network $G$, the $P({k}_{1},{k}_{2})$ is defined as the  joint probability of  finding a node with degree ${k}_{1}$  linked to a node with degree ${k}_{2}$.  $P_{r}({k}_{1},{k}_{2})$ is defined as  joint probability of the null model ${G}^{'}$ of the network $G$. The null model ${G}^{'}$ is generated by randomly rewiring all links of $G$,  while keeping the degree distribution.  We compare the correlation feature of our fractal HADGM network ($a=1,b=0.5$) with non-fractal DGM network ($e=1$) and the  World Wide Web (real-world network from Pajek Database[19]. The results are showed in Fig.4. The plots of ${R}_{HADGM}({k}_{1},{k}_{2})/{R}_{DGM}({k}_{1},{k}_{2})$ and ${R}_{HADGM}({k}_{1},{k}_{2})/{R}_{WWW}({k}_{1},{k}_{2})$ show fractal HADGM network ($a=1, b=0.5$)  have  stronger hub attraction behaviors than fractal the WWW network and non-fractal DGM network($e=1$).


\section{Assortativity and Fractality}

Relative research supports our statement is the optimization model proposed by Zheng \textit{et al}, which can produce fractal scale-free network with hubs aggregation together[18]. This model has two optimization objectives: minimizing the summation of the degrees of the nodes and maximizing the summation of the degrees of the edges. And with constraint conditions that both the $\bar{\ell}$ (average shortest path)  and  \textit{xmin}  (minimum degree of nodes throughout the entire network) are set as non-negative constants. The edge degree between node \textit{i} and node \textit{j} is defined as: ${D}_{ij}={{k}_{i}}^{m}{{k}_{j}}^{n}$, where ${k}_{i}$ and  ${k}_{j}$ are degree of node \textit{i} and node \textit{j} and \textit{m}, \textit{n} are non-negative constants.  The optimization model can produce fractal scale-free network by stretching the parameter $\bar{\ell}$. Here, we choose fractal and scale-free network generated by optimization model for comparison with parameters: node number $N=1500$, $xmin=2$, $m=0$, $n=1$, $\bar{\ell}=40$.

\begin{table}[h]
\centering  
\caption{Fractality and Disassortativity}\label{table1}
\begin{tabular}{lccccc}
\hline
\textbf{Network} & \textit{\textbf{N}} & \textbf{\textit{r}}& \textbf{Fractality} & \textbf{Disassortativity}\\
\hline
DGM $e=1$ &781245&-0.0347&NO&YES\\
HADGM $a=1, b=0.5$&784325&-0.0344&YES&YES for $k<100$\\
Optimization Network&1500 &0.8354&YES&NO\\
The Internet &22963&-0.198&NO&YES\\\hline
\hline
\end{tabular}
\end{table}
The assortativity can be measured by Pearson correlation coefficient $r$ [13], which can be rewrote as:
\begin{equation}
r=\frac{{M}^{-1}\sum_{i}^{}{j}_{i}{k}_{i}-{\left[ {M}^{-1}\sum_{i}^{}\frac{1}{2}({j}_{i}+{k}_{i})\right]}^2}
{{M}^{-1}\sum_{i}^{}\frac{1}{2}({{j}_{i}}^2+{{k}_{i}}^2)-{\left[ {M}^{-1}\sum_{i}^{}\frac{1}{2}{({j}_{i}+{k}_{i})}^2\right]}} ,
\label{pearson}
 \end{equation}
where ${j}_{i}$, ${k}_{i}$ are the degrees of the nodes at both sides of the $i$th edge, with $i = 1, . . .,M$, and $M$ is the total number of edges in a network. The positive or negative of $r$ related to assortative or disassortative mixing. As shown in Table(\ref{table1}), the DGM network($e=1$) is non-fractal and disassortative,  HADGM network ($a=1, b=0.5$) is  fractal and disassortative, the optimization network is fractal and assortative, and real network the Internet of AS-level[20] is non-fractal and disassortative.

\begin{figure*}[h]
\begin{minipage}[t]{0.5\linewidth}
\centering
\subfigure[]{
\includegraphics[width=3.in]{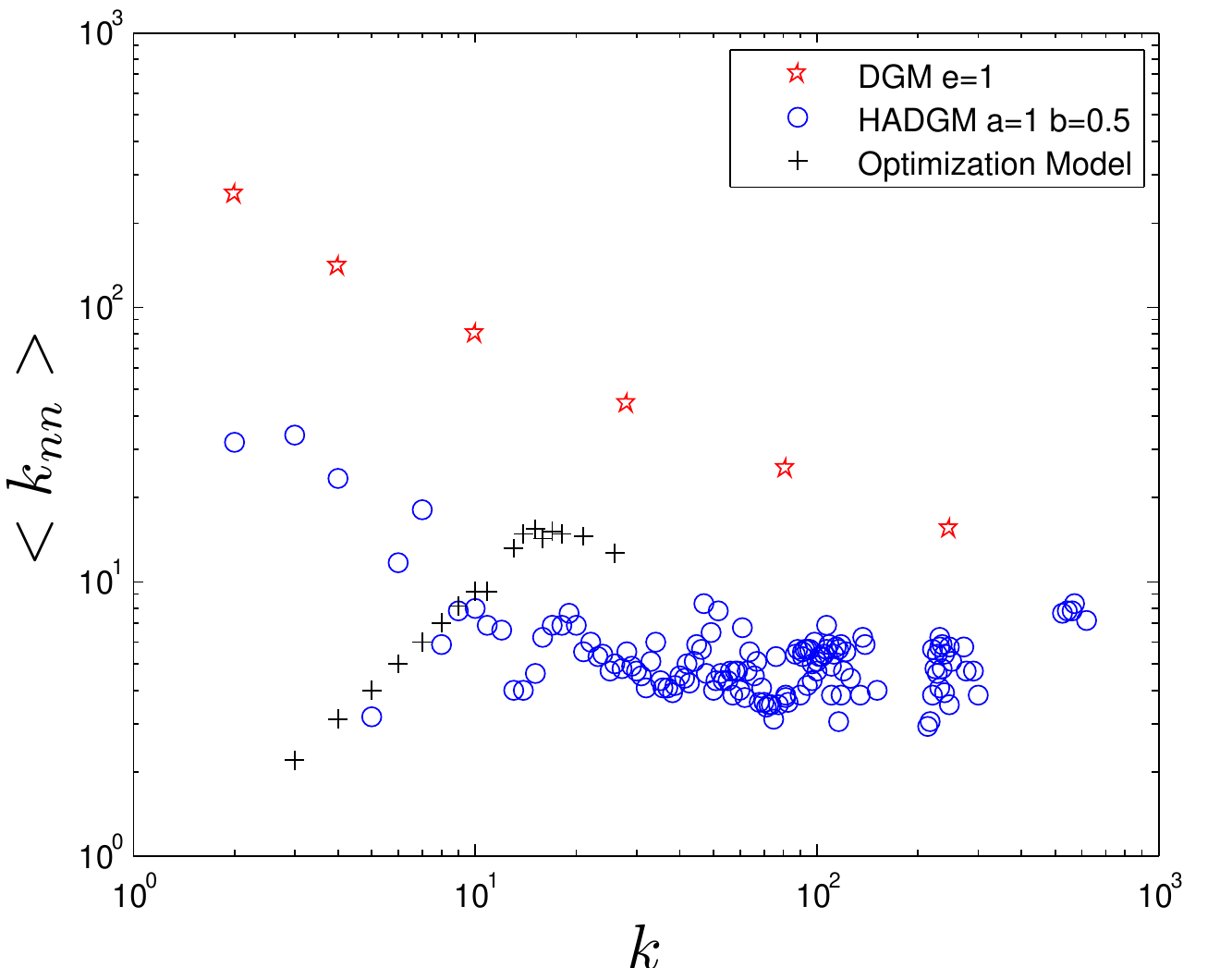}}

\label{fig:side:a}
\end{minipage}%
\begin{minipage}[t]{0.5\linewidth}
\centering
\subfigure[]{
\includegraphics[width=2.8in]{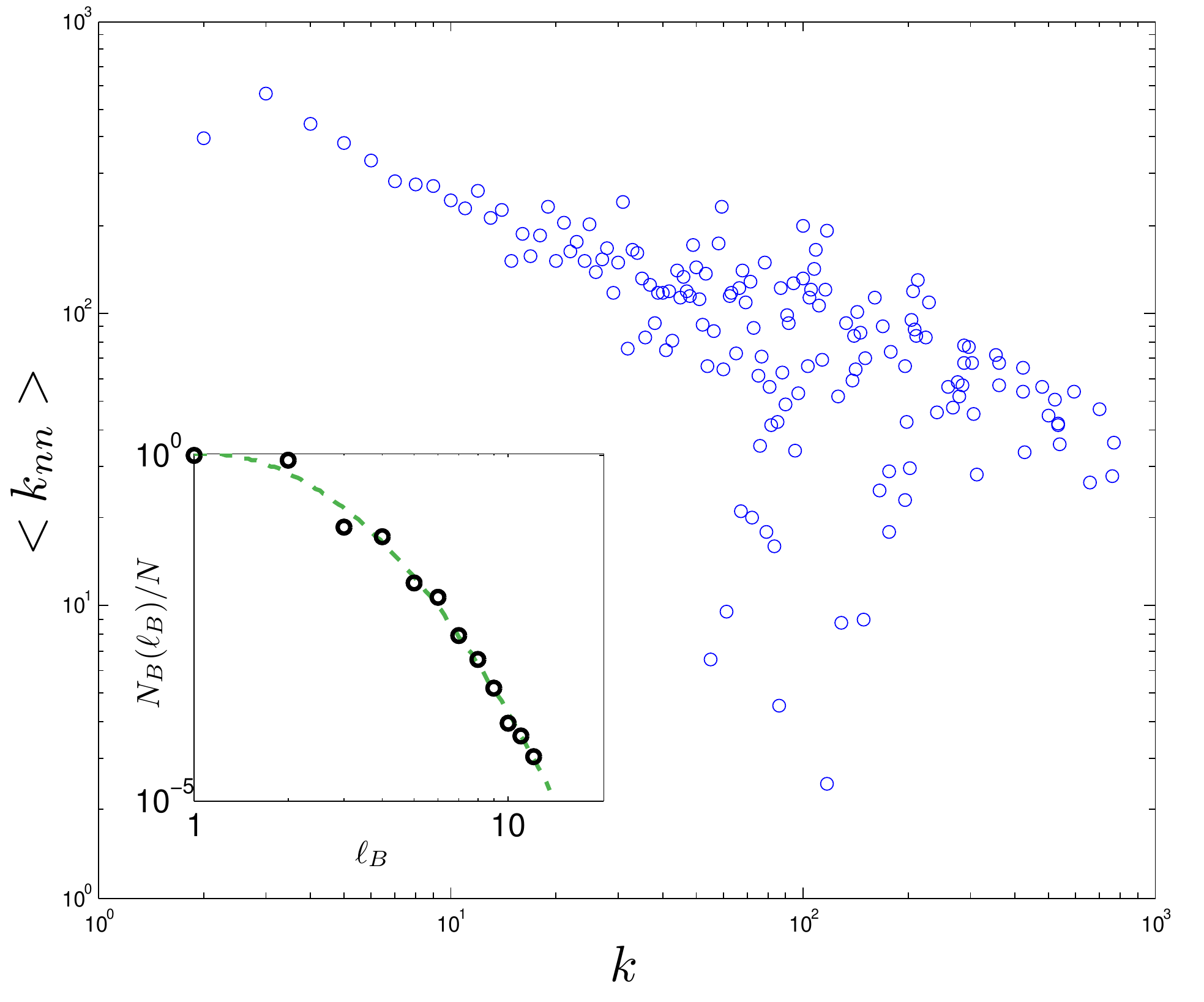}}
\end{minipage}
\caption{ Neighbor connectivity of  networks \textbf{(a)}, Neighbor connectivity of the DGM network($e=1$)(red star), the HADGM network (blue circle) and the optimization fractal network (black cross) \textbf{(b)}, Neighbor connectivity of the Internet. The inset shows the non-fractal of the Internet. }
\label{optmodel}
\end{figure*}

   We also analyze the average degree of neighbors of a node with degree $k$ namely $<{k}_{nn}>$ of above networks. $<{k}_{nn}>$ is defined as $<{k}_{nn}>=\sum\limits_{{k}^{'}} {k}^{'}P({k}^{'}|k)$, where $P({k}^{'}|k)$ is the conditional probability of an edge of node with degree $k$ connect to a node with degree ${k}^{'}$[12]. The network is assortative mixing if the slope of $k$ vs $<{k}_{nn}>$ curve is positive, which means high degree nodes tend to connect with high degree nodes, and is  disassortative mixing if the slope of $k$ vs $<{k}_{nn}>$ curve is negative[13].

   As Fig.\ref{optmodel}(a) showed,  the fractal network produced by optimization model is  assortative. And non-fractal DGM network($e=1$) is disassortative. It also shows the fractal HADGM network ($a=1, b=0.5$) is disassortative within low degrees range $k<100$, while assortative within high degrees range.  The Internet shows obvious disassortative and non-fractal property as shown in  Fig.\ref{optmodel}(b).  Therefore, the results of the DGM network ($e=1$) , optimization model and the Internet are served as counter examples of the conjecture that self-similar scale-free networks are disassortative. Furthermore, the HADGM network ($a=1, b=0.5$) has stronger correlation than DMG network  ($e=1$), yet the DMG network ($e=1$) is non-fractal and HADGM network ($a=1, b=0.5$) is fractal. Our results show the fractality property is  independent of the assortative mixing feature.

\section{Conclusion and Discussion}

 In this paper we proposed a new dynamical growth model, which can produce fractal networks with hub attraction behaviors. The model is designed by applying flexible probability of cross-boxes links mechanism and inside box link-growth method.  More general models can also be proposed by applying different function of flexible probability $e$ or other kinds of inside link-growth mechanism.

 Observing equations (\ref{nodedeg}) and (\ref{fradimension}), the fractality property requires the diameter $\tilde{L}(t)$ of networks exponential growth through time evolution rather than linear growth. Therefore, we believe there is a structural equilibrium in fractal networks to keep the diameter $\tilde{L}(t)$  within a certain range. According to the topological structure of complex networks, we could divide a network into several layers as mentioned in[[15]. The most connected hub is consider as the center, other nodes are allocated in different layers depending on the distance to the center node. Hub attraction and booming growth in boundary are a pair of opposing force to maintain the structural equilibrium in fractal network. Hub attraction will lead to  $\tilde{L}(t)$ to drop dramatically, while booming growth in boundary will result in $\tilde{L}(t)$ increasing sharply .

\appendix
\section{Appendix}
According to the HADGM model, at time step $t+1$ we generate $mk(t)$ new nodes for  each node with degree $k(t)$ at time step $t$. Therefore, we have:
\begin{equation}
\tilde{N}(t+1)= \tilde{N}(t)+2m\tilde{K}(t),
\label{nodenumber1}
 \end{equation}
 where $\tilde{K}(t)$ is total number of links at time step $t$.
At time step $t+1$, in the flexible probability growth phase we have $(2m+1)\tilde{K}(t)$ edges in network, then in the link-growth inside box phase we add $2\tilde{K}(t)$ edges. Thus, we have:
  \begin{equation}
\tilde{K}(t+1)= (2m+3)\tilde{K}(t).
\label{nodedeg1}
 \end{equation}
Combine equation(\ref{nodenumber1}) and equation(\ref{nodedeg1}), we have:
\begin{equation}
  \tilde{N}(t)= \tilde{N}(0)+m \tilde{K}(0)\frac{(2m+3)^t-1}{m+1}.
\label{nodenumber2}
 \end{equation}
 Due to $\tilde{N}(t)\gg\tilde{N}(0)$, we can acquire $\tilde{N}(t)\thickapprox (2m+3)\tilde{N}(t-1)$ for $t>1$.

 Considering the node degrees growth with time evolution, for Model I, alone we have  $\tilde{k}(t)= (m+1)\tilde{k}(t-1)$ and for Model II alone, we have  $\tilde{k}(t)= m\tilde{k}(t-1)$. Therefore, for probability $\bar{e}$ to combine Model I and Model II, we have $ \tilde{k}(t)= (m+\bar{e})\tilde{k}(t-1)$.

At last, we look at the diameter growth with time evolution. For Model I alone, we have $\tilde{L}(t)= \tilde{L}(t-1)+2$ and for Model II alone, we have  $\tilde{L}(t)= 3\tilde{L}(t-1)$. Therefore, for probability $\bar{e}$ to combine Model I and Model II, we have $ \tilde{L}(t)= (3-2\bar{e})\tilde{L}(t-1)+2\bar{e}$.

\renewcommand{\thefootnote}

\Acknowledgements{}

\normalsize \vskip0.3in\parskip=0mm \baselineskip 18pt
\renewcommand{\baselinestretch}{1.1}\footnotesize\parindent=4mm\bahao

\vskip0.1in \noindent {\normalsize \bf References}
\vskip0.1in\parskip=0mm

\REF{1\ }
B.~B. Mandelbrot, {\em The fractal geometry of nature}.
\newblock Macmillan, 1983.

\REF{2\ }
H.-O. Peitgen, H.~J{\"u}rgens, and D.~Saupe, {\em Chaos and fractals - new
  frontiers of science (2. ed.)}.
\newblock Springer, 2004.

\REF{3\ }
M.~E. Newman, ``The structure and function of complex networks,'' {\em SIAM
  review}, vol.~45, no.~2, pp.~167--256, 2003.

\REF{4\ }
R.~A. Albert-L\'{a}szl\'{o}~Barab\'{a}si, ``Emergence of scaling in random
  networks,'' {\em Science}, vol.~286, pp.~509--512, Oct. 1999.

\REF{5\ }
A.-L.~B. R\'{e}ka~Albert, ``Statistical mechanics of complex networks,'' {\em
  Rev. Mod. Phys.}, vol.~74, pp.~47--97, Jan. 2002.

\REF{6\ }
C.~Song, S.~Havlin, and H.~A. Makse, ``Self-similarity of complex networks,''
  {\em Nature}, vol.~433, pp.~392--395, Jan. 2005.

\REF{7\ }
C.~Song, S.~Havlin, and H.~A. Makse, ``Origins of fractality in the growth of
  complex networks,'' {\em Nat Phys}, vol.~2, pp.~275--281, Apr. 2006.

\REF{8\ }
C.~Song, L.~K. Gallos, S.~Havlin, and H.~A. Makse, ``How to calculate the
  fractal dimension of a complex network: the box covering algorithm,'' {\em
  Journal of Statistical Mechanics: Theory and Experiment}, vol.~2007, no.~03,
  pp.~P03006--, 2007.

\REF{9\ }
J.~Kim, K.~Goh, B.~Kahng, and D.~Kim, ``Fractality and self-similarity in
  scale-free networks,'' {\em New Journal of Physics}, vol.~9, no.~6, p.~177,
  2007.

\REF{10\ }
W.-X. Zhou, Z.-Q. Jiang, and D.~Sornette, ``Exploring self-similarity of
  complex cellular networks: The edge-covering method with simulated annealing
  and log-periodic sampling,'' {\em Physica A: Statistical Mechanics and its
  Applications}, vol.~375, no.~2, pp.~741--752, 2007.

\REF{11\ }
C.~M. Schneider, T.~A. Kesselring, J.~S. Andrade~Jr, and H.~J. Herrmann,
  ``Box-covering algorithm for fractal dimension of complex networks,'' {\em
  Physical Review E}, vol.~86, no.~1, p.~016707, 2012.

\REF{12\ }
R.~Pastor-Satorras, A.~V{\'a}zquez, and A.~Vespignani, ``Dynamical and
  correlation properties of the internet,'' {\em Physical review letters},
  vol.~87, no.~25, p.~258701, 2001.

\REF{13\ }
M.~E. Newman, ``Assortative mixing in networks,'' {\em Physical review
  letters}, vol.~89, no.~20, p.~208701, 2002.

\REF{14\ }
S.-H. Yook, F.~Radicchi, and H.~Meyer-Ortmanns, ``Self-similar scale-free
  networks and disassortativity,'' {\em Physical Review E}, vol.~72, no.~4,
  p.~045105, 2005.

\REF{15\ }
J.~Shao, S.~V. Buldyrev, R.~Cohen, M.~Kitsak, S.~Havlin, and H.~E. Stanley,
  ``Fractal boundaries of complex networks,'' {\em EPL (Europhysics Letters)},
  vol.~84, no.~4, pp.~48004--, 2008.

\REF{16\ }
K.-I. Goh, G.~Salvi, B.~Kahng, and D.~Kim, ``Skeleton and fractal scaling in
  complex networks,'' {\em Phys. Rev. Lett.}, vol.~96, pp.~018701--, Jan. 2006.

\REF{17\ }
S.~Maslov and K.~Sneppen, ``Specificity and stability in topology of protein
  networks,'' {\em Science}, vol.~296, no.~5569, pp.~910--913, 2002.

\REF{18\ }
B.~Zheng, H.~Wu, J.~Qin, W.~Du, J.~Wang, and D.~Li, ``A simple model clarifies
  the complicated relationships of complex networks,'' {\em CoRR},
  vol.~abs/1210.3121, 2012.

\REF{19\ }
V.~Batagelj, A.~Mrvar, Pajek datasets,
  http://vlado.fmf.uni-lj.si/pub/networks/data/ (2006).
\REF{20\ }
L.~Wang, ``Internet topology collection.'' http://irl.cs.ucla.edu/topology/.

\end{document}